\documentclass[aps,prl,preprint,12pt,altaffilletter,showpacs]{revtex4}
\usepackage{graphicx}
\usepackage{amssymb}
\usepackage{amsmath} 
\usepackage{hyperref}
\usepackage{textcomp}
\usepackage{psfrag}
\usepackage{setspace}
\usepackage{url}
\usepackage{bm}
\usepackage{subfigure}
\begin{document}
\title{Hysteresis and Lubrication in Shear Thickening of Cornstarch
Suspensions}
\author{Clarence~E.~Chu}
\author{Joel~A.~Groman}
\author{Hannah~L.~Sieber} 
\author{James~G.~Miller}
\affiliation{Department of Physics, Washington University, St.~Louis,
Mo.~63130}
\author{Ruth~J.~Okamoto}
\affiliation{Department of Mechanical Engineering and Materials Science, 
Washington University, St.~Louis, Mo.~63130}
\author{Jonathan I.~Katz$^*$}
\affiliation{Department of Physics, Washington University, St.~Louis,
Mo.~63130}
\affiliation{McDonnell Center for the Space Sciences, Washington
University, St.~Louis, Mo.~63130}
\email{katz@wuphys.wustl.edu}
\date{\today}
\begin{abstract}
Aqueous and brine suspensions of cornstarch show striking discontinuous
shear thickening.  We have found that a suspension shear-thickened
throughout may remain in the jammed thickened state as the strain rate is
reduced, but an unjamming front may propagate from any unjammed regions.
Transient shear thickening is observed at strain rates below the thickening
threshold, and above it the stress fluctuates.  The jammed shear-thickened
state may persist to low strain rate, with stresses resembling sliding
friction and effective viscosity inversely proportional to the strain rate.
At the thickening threshold fluid pressure depins the suspension's contact
lines on solid boundaries, so that it slides, shears, dilates and jams.  In
oil suspensions lubrication and complete wetting of confining surfaces
eliminate contact line forces and prevent jamming and shear thickening, as
does addition of immiscible liquid surfactant to brine suspensions.  Starch
suspensions in glycerin-water solutions, viscous but incompletely wetting,
have intermediate properties.

\end{abstract}
\maketitle
\section{Introduction} 
Aqueous and brine suspensions of cornstarch have long been known \cite{WH31,
FR38,B89,F08,F12} to show the dramatic property of discontinuous shear
thickening (DST).  Cornstarch is a complex biogenic substance consisting of 
polydisperse grains with mean diameter $\approx 14\,\mu$ and blocky
irregular shapes.  The ready observation of DST in cornstarch suspensions
over a broad range of starch fractions and conditions illustrates the
robustness and universality of the phenomenon.

Shear thickening interferes with some industrial processes \cite{B89} but
may be useful to absorb mechanical energy \cite{BJ09,WCR09} or to suppress
hydrodynamic instability \cite{BLMK11}.  Thickened starch suspensions show
quasi-solid behavior: rapid stirring produces tensile fracture and uncovers
the bottom of a shallow container.  In contrast, starch suspensions in
nonpolar fluids, such as oils, benzene and CCl$_4$, are unremarkable shear
thinning fluids \cite{WH31,PJ41}.  An explanation of shear thickening of
starch suspensions in water and brine must also explain its absence in
starch suspensions in oil.

Cornstarch occupies an intermediate regime between colloids and macroscopic
particles.  Interactions \cite{H74} that lead to non-Newtonian properties of
colloidal suspensions are insignificant.  Starch grains are small enough to
form suspensions, yet large enough that these suspensions are non-Brownian:
the P\'eclet number at the DST threshold ${\rm Pe} \equiv {a^2 {\dot \gamma}
\over D} = {6 \pi \eta_f a^3 {\dot \gamma} \over k_B T} \gtrsim$ 2000,
where $a \approx 7\,\mu$ is the mean particle radius, the strain rate ${\dot
\gamma} = {\dot \gamma}_c \gtrsim$ 2/s where ${\dot \gamma}_c$ is the
critical strain rate for shear thickening, $\eta_f$ the dynamic viscosity of
the solvent and $D$ the Brownian diffusivity.

\section{Hysteresis}

\begin{figure}[h!]
\centering
\subfigure[\ Cone]{
\includegraphics[width=0.45\textwidth]{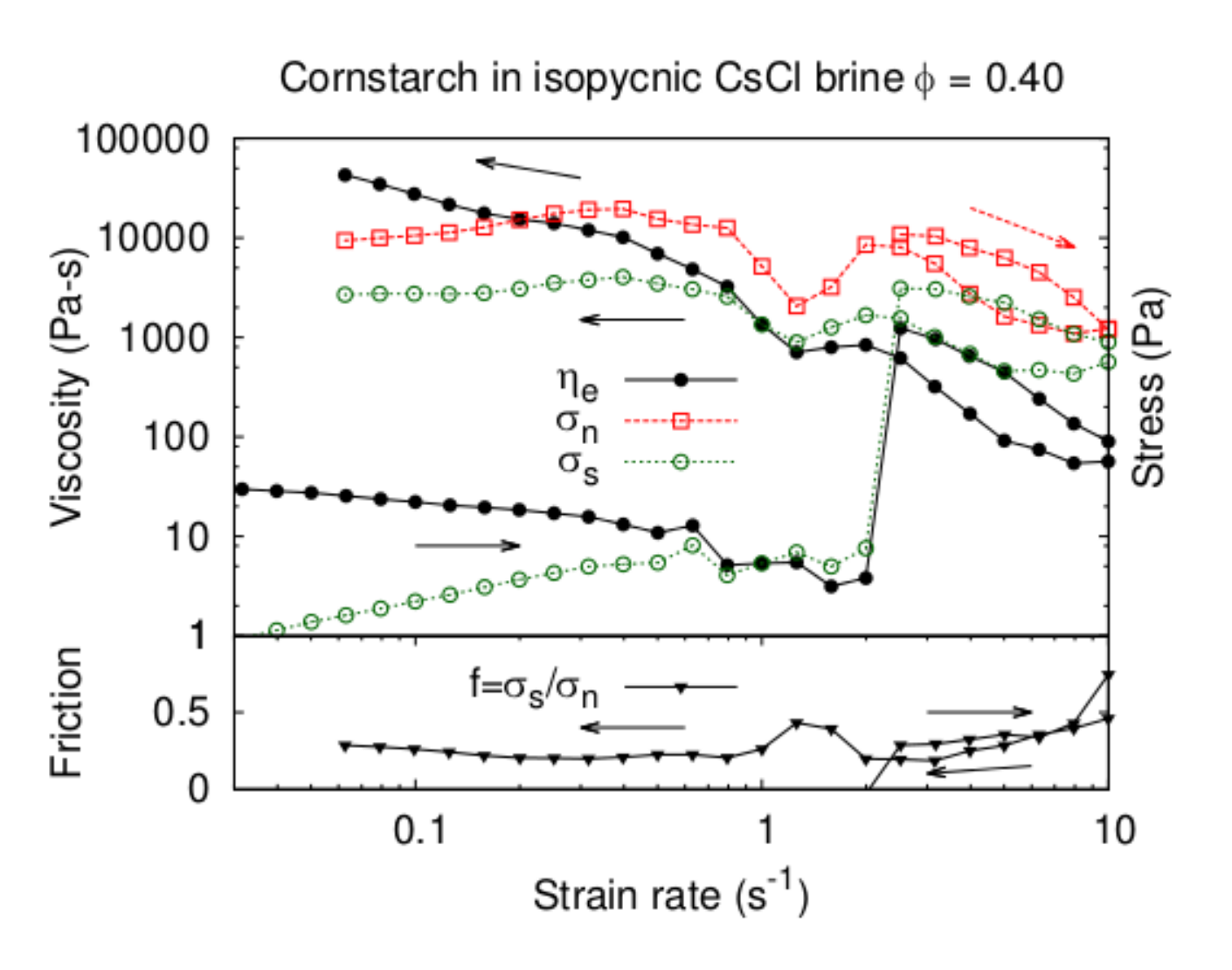}
\label{brinecone}}
\subfigure[\ Parallel plates]{
\includegraphics[width=0.45\textwidth]{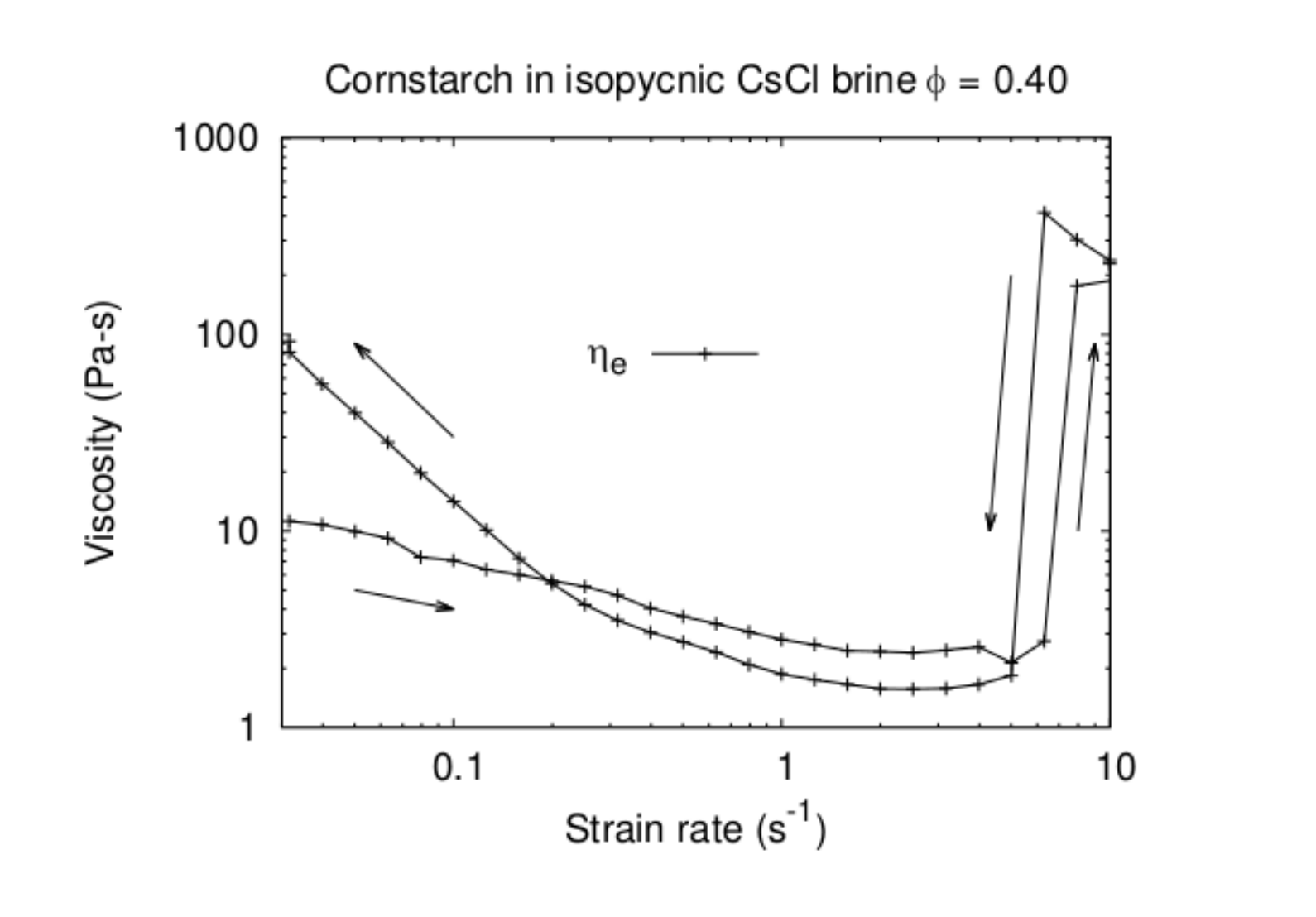}
\label{brineplate}}
\caption{\label{brine}Rheology of suspensions of cornstarch (starch fraction
$\phi = 0.40$) in isopycnic (53.5\%) CsCl aqueous brine.  The strain rate
was increased from 0.03/s to 10/s, and then decreased (directions shown with
arrows) in discrete steps; each point represents a 60 s measurement at
constant strain rate.  Lines guide the eye between data points.  (a)
Homogeneous strain rate in conical (2$^\circ$) geometry.  The normal stress
$\sigma_n$ is small and negative ($\simeq -200\,$Pa, nearly independent of
strain rate; not shown), attributable to meniscus tension, until DST, but
then becomes large, positive and very roughly independent of strain rate.
Even low strain rates are sufficient to maintain the starch grains in their
jammed state.  The shear stress $\sigma_s$ follows a sigmoidal curve
\cite{WC14} as the strain rate is increased, but remains high, like
$\sigma_n$, even as the strain rate is reduced below ${\dot \gamma}_c$.
Their ratio, an effective sliding friction coefficient \cite{B10}, $f \equiv
\sigma_s/\sigma_n \approx 0.3$.  The effective viscosity $\eta_e \equiv
\sigma_s/{\dot \gamma} \propto {\dot \gamma}^{-1}$ in the shear-thickened
regime.  (b) In parallel plate geometry (0.7 mm) the strain rate varies from
zero on the axis to its maximal value ($\dot \gamma$, shown) at the
periphery.  Hysteresis is minimal; the suspension unjams as the strain rate
is reduced.  The larger value of ${\dot \gamma}_c$ may also be attributed to
an unjamming front propagating from regions of lower $\dot \gamma$ near the
axis.  [Color on-line.]}
\end{figure}

The richness of the rheological properties of cornstarch suspensions
includes remarkable hysteresis \cite{HAC01,L03,KSM13,WC14}.  Fig.~\ref{brine}
shows the behavior of suspensions of cornstarch in density-matched CsCl
brine.  The strain rate was first stepped up from low values into the
shear-thickened regime, and then stepped down.  When shear is imposed
between a rotating cone and a static flat plate the strain rate is
homogeneous.  Hysteresis is evident; once the suspension has jammed in the
shear-thickened state, even a low strain rate is sufficient to keep it
jammed (Fig.~\ref{brinecone}).  We infer (but cannot directly observe) that
the entire suspension jams.  The normal stress imposed by the rheometer to
maintain a constant gap width is small and negative \cite{L05} in the
unthickened state but large, positive, and roughly constant once the
suspension has thickened, even as $\dot \gamma$ is stepped down below ${\dot
\gamma}_c$.  We attribute this to the large normal stress and static
friction between grains maintaining the jammed state against the shear
stress.

In contrast, in parallel plate geometry the suspension thickened at ${\dot
\gamma}_c \approx$ 6--8/s and unthickened when $\dot \gamma$ was reduced
slightly below this value, as shown in Fig.~\ref{brineplate}.  This suggests
that an unjamming front propagated from unjammed suspension near the
rotation axis where ${\dot \gamma} < {\dot \gamma}_c$ into jammed material,
as granules on the front became free to reorient themselves and force chains
were disrupted.  Analogous behavior is found for suspensions of BiOCl
\cite{BBS02} in which contact with fluid material initiates a liquifaction
front that propagates into jammed quasi-solid material, and similar
phenomena may be involved in mudslides.



\begin{figure}[h!]
\centering
\subfigure[\ Cone]{
\includegraphics[width=0.45\textwidth]{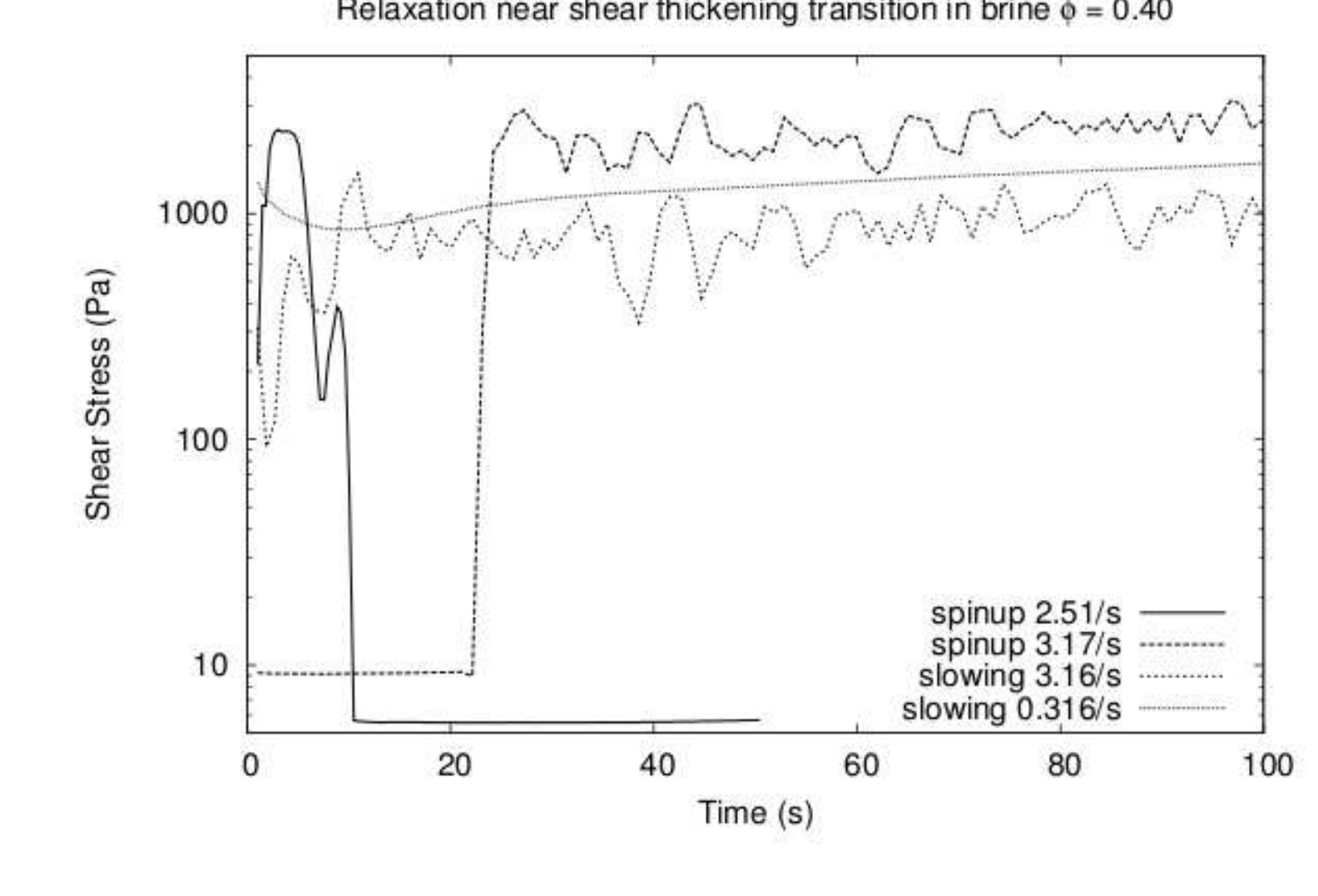}
\label{fluctcone}}
\subfigure[\ Parallel plates]{
\includegraphics[width=0.45\textwidth]{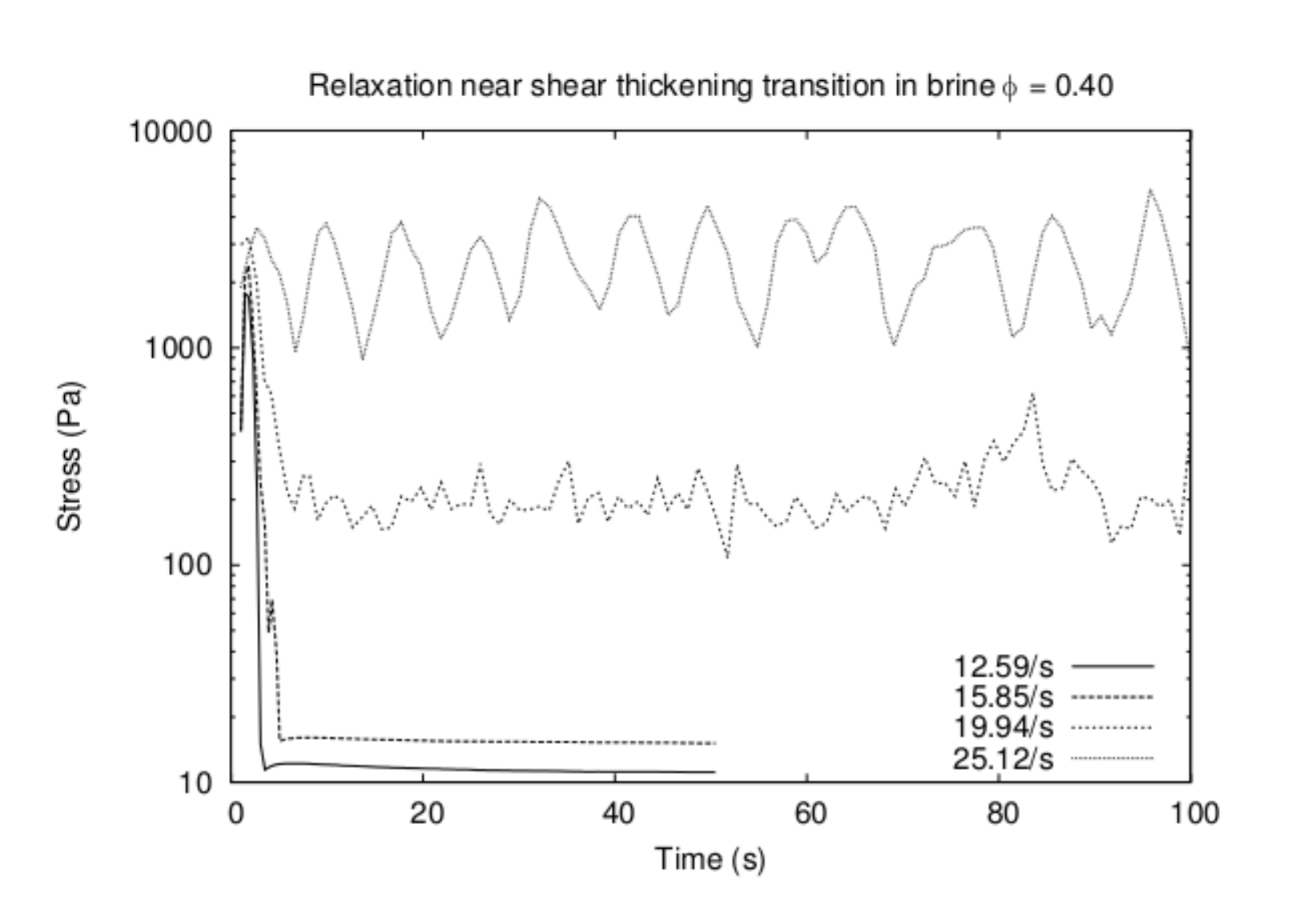}
\label{fluctplate}}
\caption{\label{fluct}Variations of shear stress $\sigma_s$ with time at
constant strain rate $\dot \gamma$ after stepping up (spinup) or down
(slowing); (a) In conical geometry ($2^\circ$) $\sigma_s$ fluctuates in the
jammed state after DST at 2.51/s $ < {\dot \gamma}_c < $ 3.17/s; similar
behavior was observed as $\dot \gamma$ was increased to 10/s, and
subsequently decreased.  After slowing to $\dot \gamma \ll {\dot \gamma}_c$,
$\sigma_s$ was roughly unchanged but steady (${\dot \gamma} = $ 0.316/s
shown). (b) In parallel plate geometry ($\dot \gamma$ at rim, 0.5 mm) for
${\dot \gamma} < $ 10/s $\sigma_s$ relaxes within 1 s to a new steady value
with no indication of transient behavior (not shown).  For slightly higher
${\dot \gamma} \lessapprox {\dot \gamma}_c \approx 20$/s there is transient
thickening, followed by relaxation over a few seconds into a less viscous
quicksand-like state.  For ${\dot \gamma} \gtrapprox {\dot \gamma}_c$ the
suspension shear thickens as granules jam and unjam and $\sigma_s$ 
fluctuates with large amplitude rather than settling to a steady value.  
This behavior continues to strain rates several times higher than ${\dot
\gamma}_c$ (as high as it has been possible to shear the suspension without
expelling it centrifugally).}
\end{figure}

Even at steady strain rates, hysteresis is observed near the DST threshold
(Fig.~\ref{fluct}).  For $\dot \gamma$ slightly below ${\dot \gamma}_c$ (but
not for yet smaller $\dot \gamma$) each increment in $\dot \gamma$ produces
an initial thickening followed by rapid relaxation into a steady and less
viscous state.  This may be explained as the disruption of transient force
chains \cite{CWBC98} by the shear flow when the stress is insufficient to
maintain frictional contacts within force chains and between them and
confining surfaces.

At and above the DST threshold (which is steep but not truly discontinuous;
Fig.~\ref{fluctplate}) $\sigma_s$ fluctuates with large amplitude,
suggesting a quasi-solid jammed suspension, maintained by a larger normal
stress $\sigma_n$, that intermittently slips, fractures, or partially jams
and unjams.  These fluctuations might suggest stick-slip friction, but this
is not consistent with the steady $\sigma_s$ when $\dot \gamma$ is reduced
but the suspension remains jammed in conical geometry (${\dot \gamma} =
0.316/$s in Fig.~\ref{fluctcone}).  The fact that the fractional fluctuation
amplitude is greater for parallel plates also suggests a contribution from
fluctuation of a jammed/unjammed boundary at small radius, rather than it
being entirely a stick-slip phenomenon independent of the specific geometry.


This fluctuating stress in the DST regime resembles that observed for
colloids \cite{H09} at much higher strain rates.  In experiments with larger
(5.8$\,\mu$) non-Brownian particles (expected to be in the same regime as
cornstarch), in which stress was the control parameter, a fluctuating $\dot
\gamma$ was observed \cite{L14}.  However, this may be explained by a
multivalued ${\dot \gamma}(\sigma_s)$ and mechanical relaxation between its
two values rather than by intrinsic hysteresis.

Fluctuating stress cannot be explained as a consequence of statistical
fluctuations in the number of independent force chains:  A simple estimate,
using Hertzian contact theory, shows that for a mean stress $\sigma_n$
applied over an area $A$ to a layer of thickness $h$ of grains of Young's
modulus $E$ \cite{JHMK13}, radius $a$, and radius of curvature at their
contacts $r_{curv}$, the number of force chains carrying the load 
\begin{equation}
n_{chain} \simeq {A \over a^2} \left({\sigma_n \over E}\right)^{2/5}
\left({h \over a}\right)^{3/5} \left({a \over r_{curv}}\right)^{1/5} \simeq
10^6 \left({a \over r_{curv}}\right)^{1/5} \gg 1.
\end{equation}
Even jamming on the scale $h$ would lead to $n_{block} \simeq A/h^2 \simeq
2000$ of independent blocks.  The amplitude of the measured stress
fluctuations implies that the entire jammed volume acts as a single solid
body.

\section{Surfaces and Confinement}

The values of ${\dot \gamma}_c$ and of the effective viscosity $\eta_e({\dot
\gamma}_c)$ in the shear-thickened state require explanation.  The dimensional
parameters $a$ and $\nu_f$ define a dimensional strain rate ${\dot \Gamma}
\equiv \nu_f/a^2 \approx 2 \times 10^4{\rm /s} \gg {\dot \gamma}_c$, where
$\nu_f$ is the kinematic viscosity of the solvent; substitution of the
suspension viscosity for $\nu_f$ would multiply Pe and $\nu_f/a^2$ by an
additional large factor.  No dimensionless groups exist to multiply or
divide $\dot \Gamma$ to bring it into agreement with ${\dot \gamma}_c$.  

An additional physical effect and dimensional parameter are required.  The
introduction of a surface energy $\bm{\gamma}$ permits defining a new
dimensionless parameter, the surfluidity: 
\begin{equation}
{\rm Su} \equiv {\bm{\gamma} s \over \rho \nu_f^2},
\end{equation}
where $s$ is a characteristic length, such as the particle radius $a$ or
layer thickness $h$.  Su measures the comparative importance of surface and
viscous forces and is equivalent to the Reynolds number of a partially
wetted particle or flow acted upon only by surface forces at its contact
line and by viscosity.  When two surfaces are brought together the sign of
$\bm{\gamma}$ is significant: if the fluid reduces the interfacial energy
then $\bm{\gamma} < 0$ and a lubricating layer persists, but if $\bm{\gamma}
> 0$ surface forces expel this layer, bringing the solids into dry contact.
If $s = a$ then for cornstarch in water or brine $\vert {\rm Su} \vert
\approx 200$, but in olive oil $\vert {\rm Su} \vert \approx 0.06$ (taking
$\vert \bm{\gamma} \vert = 50\,$dyne/cm, a representative value for
molecular liquids).   Surface energies are important for low viscosity
solvents, but viscosity may be sufficient to maintain lubrication by more
viscous solvents.

DST of aqueous and brine starch suspensions is attributed
\cite{F08,BJ09,B10,BJ12,BJ13} to confinement of the grains by surface
tension, in analogy to the shear thickening of larger particles confined by
solid walls.  During shear thickening the suspension has a ``dry'' or
``rough'' appearance \cite{MW58,HFC03,BJ09,BJ12,BJ13,C05} indicative of
particles pushed through the surface of the solvent as a consequence of
shear dilatancy \cite{R1885} (packed particles dilate when sheared because
they are subject to too many constraints on displacement and rotation at
unlubricated contacts to permit shear until some constraints are relieved by
dilation).  Surface tension acting on these particles explains the viscosity
in the thickened state.  It applies a confining stress \mbox{$\sigma_c =
{\rm O}(\bm{\gamma}/a) = {\rm O}(10^4$ Pa)} \cite{MW58, BJ12} to the
particles.  Unjammed, particles are individually pushed back into the body
of the fluid, but if they are jammed this confining compressive stress is
applied to the entire particle network.  This also holds (with $\bm{\gamma}$
the interfacial tension) if the suspension is immersed in an immiscible
fluid, such as a brine suspension immersed in oil \cite{BLMK11}.  
If there is no opposing momentum
supplied by the walls or bottom of a container \cite{LSZ10} then
momentum conservation requires a balancing tension in the fluid \cite{D09}.
The fluid tension may be sufficient to induce fracture or cavitation, as
observed in a vigorously stirred suspension in an open vessel.

In the Coulomb-Mohr model of the strength of solids and of granular
materials with unlubricated grain-grain static friction the shear strength
is comparable to $\sigma_c$.  The flow condition in the jammed state is that
$\sigma_s$ equal the strength, so that $\sigma_s \sim \sigma_c \sim
\sigma_n$ ($\sigma_c \sim \sigma_n$ is required by the constrained rheometer
geometry), independent of $\dot \gamma$.  
Sliding friction against confining surfaces \cite{CWBC98}, with an effective
coefficient $f$, produces shear thickening with shear stress $\sigma_s = f
\sigma_n = {\rm O} (f\bm{\gamma}/a)$ and effective viscosity $\eta_e \equiv
\sigma_s/{\dot \gamma} \propto {\dot \gamma}^{-1}$ \cite{BJ12}.
Fig.~\ref{brinecone} shows this behavior both as $\dot \gamma$ increases and
also as it decreases far below ${\dot \gamma}_c$; Hoffman \cite{H72,H82}
found a similar result for ${\dot \gamma} > {\dot \gamma}_c$.  The measured
$\sigma_n \simeq 10^4$ Pa is comparable to $\bm{\gamma}/a$ for starch
granules, as predicted.  This model assumes that the differential motion of
the rheometer surfaces is at least partly accommodated by shear (implying
dilation) of the suspension, rather than entirely by surface slip.  If it
were entirely surface slip, the unsheared suspension would not be dilated or
jammed, an inconsistency.


\section{DST Threshold}
\begin{figure}
\begin{center}
\includegraphics[width=1.5in]{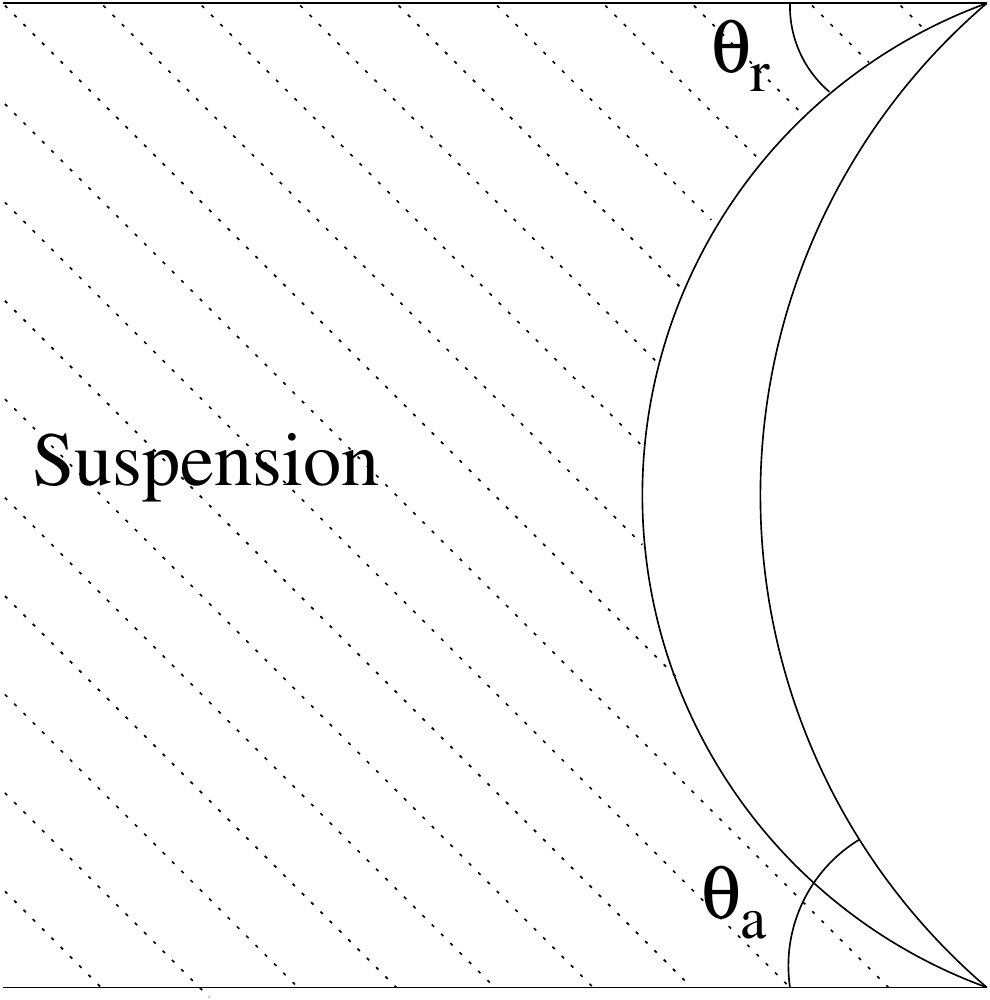}
\end{center}
\caption{\label{gap}Suspension confined between horizontal base plate and
rotating parallel plate or shallow cone in a rheometer.  Incipient dilation
applies a force determined by the shear stress in the unjammed
(shear-thinned) state to the three-phase contact line.  If this force is
sufficient to advance the contact line then the granules dilate and jam
against the confining surfaces, producing discontinuous shear thickening.}
\end{figure}

Surface tension may also explain the strain rate threshold ${\dot \gamma}_c$
of DST.  A suspension that is free to expand does not undergo DST;
confinement is required \cite{F08,BJ09,B10,BJ12,BJ13}.  If at the edge of a
gap of width $h$ there is a free surface rather than additional fluid, as
shown in Fig.~\ref{gap}, there is a confining meniscus stress on both
granules and fluid ${\rm O} (\bm{\gamma}/h) = {\rm O} (70$ Pa), where $h
\approx 0.07$ cm in most experiments, (the surface tension of CsCl brine is
close to that of water \cite{A92}).  This is also the source of the negative
$\sigma_n$ in the unstiffened state.  Although much less than the internal
stresses in the dilatant sheared state (because $h \gg a$), this stress is
present even without strain.

A force chain \cite{CWBC98} carrying a force $F$ and making an angle
$\theta$ to the normal and an azimuthal angle $\phi$ with respect to the
flow direction applies a normal force $F \cos{\theta}$ and a tangential
force $F \sin{\theta} \cos{\phi}$.  Averaging over an isotropic distribution
of force chains in $0 \le \theta \le \pi/2$ and $-\pi/2 \le \phi \le \pi/2$
(force chains cannot support tension) yields $\sigma_{s,chain} =
\sigma_{n,chain}$.  Sliding friction on the confining surfaces with
coefficient $f$ implies $\sigma_{s,chain} = \sigma_s = f \sigma_n$.  Some of
the normal stress must be carried by a fluid pressure 
\begin{equation}
\label{fluidp}
P = \sigma_n - \sigma_{n,chain} = \sigma_s \left({1 \over f} -1\right)
\simeq 2 \sigma_s
\end{equation}
for $f \simeq 0.3$, as estimated in the fully shear-thickened state
(Fig.~\ref{brinecone}).

The maximum fluid pressure that can be sustained without moving the meniscus
\begin{equation}
P_{max} = {2 \bm{\gamma} \over h} (\cos\theta_r - \cos\theta_a).
\label{pmax}
\end{equation}
The receding and advancing contact angles $\theta_r$ and $\theta_a$ depend
empirically on the surfaces and their condition, but in Fig.~\ref{brine} $h
= 0.07\,$cm, the difference in cosines $\approx 0.15$ \cite{L02} and
$P_{max} \approx 20\,$Pa.  This value is consistent with the measured 
$\sigma_s \simeq 10\,$Pa and inferred fluid $P \simeq 20\,$Pa
(Eq.~\ref{fluidp}) in the unthickened state at the threshold of DST shown
in Fig.~\ref{brine} and in \cite{F08}.  Hoffman \cite{H74} found this stress
to be independent of $\nu_f$ over three orders of magnitude, indicating
that the thickening threshold is determined by surface tension and
lubrication rather than by bulk properties of the solvent.

DST occurs in a steady flow with a free surface if (and only if) shear
produces a dilational stress that moves the contact line between suspension,
air and confining surface.  The strain rate ${\dot \gamma}_c$ at the DST
threshold is determined by $P_{max}$ and the suspension viscosity (not that
of the solvent) in the unstiffened state; stress and $\dot \gamma$,
rather than $\nu_f$, are the controlling parameters.  At lower strain rates
and stresses the meniscus deforms but the contact line does not move; the
suspension does not slide on the confining surfaces and the granules do not
jam.  Once $P > P_{max}$ the suspension slides along the confining surfaces
to accommodate dilation of the granules.  With sufficient friction, dilation
jams the granules against these surfaces \cite{CWBC98,SMMD13} and triggers
DST.  This is consistent with the observation \cite{L14} that wall slip is
dramatically reduced or disappears when a suspension enters the dilatant
regime. 

\section{Other Suspending Fluids}

\begin{figure}
\centering
\includegraphics[width=3in]{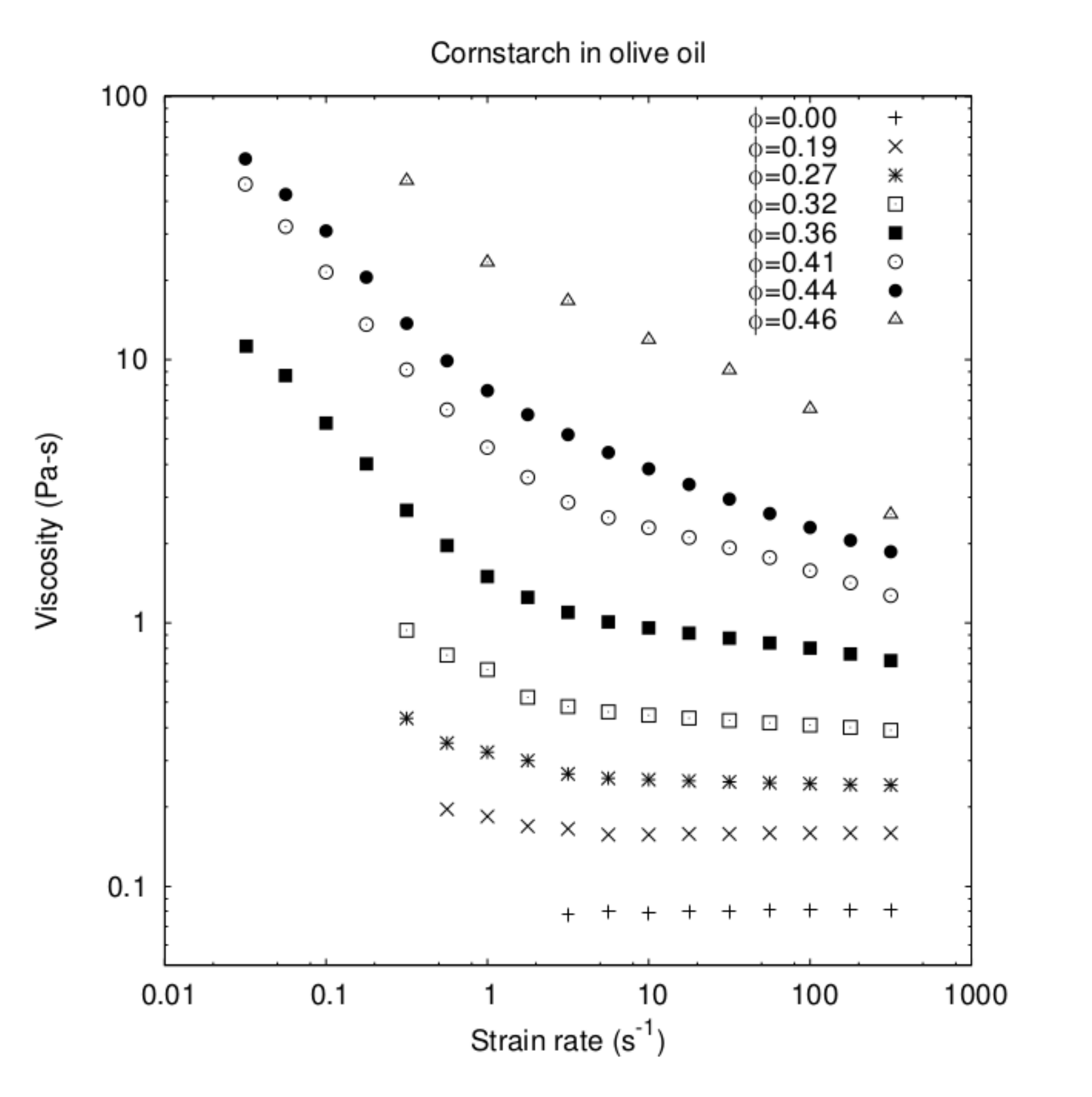}
\caption{\label{oilfig}Rheology of olive oil suspensions of cornstarch
measured in a rotating $2^\circ$ conical gap with 2 cm radius.  Each point
represents a steady state at constant strain rate.  It was not possible to
obtain reproducible data at very low strain rates for the most concentrated
suspensions and or low torques (low strain rates for dilute suspensions).
Shear thinning is evident, but not shear thickening.  The normal stress
$\sigma_n$ remains small and negative, as for brine suspensions in the
unthickened regime.  One run at a starch volume fraction $\phi = 0.46$
provided evidence of a finite yield stress with divergent viscosity for
${\dot \gamma} < 0.01$/s, and such concentrated mixtures may resemble moist
pastes rather than suspensions, but there was no evidence of a yield stress
for $\phi \le 0.44$.}
\end{figure}

We test these hypotheses by studying suspensions in fluids with different
properties.  Fig.~\ref{oilfig} shows the rheology of suspensions of
cornstarch in olive oil.  It is not possible to match the density of oil
to that of starch, but because the viscosity of olive oil is about 80 times
that of water or brine \cite{CRC68,NST95} and these suspensions are
concentrated \cite{C00}, sedimentation is slow.  There is no shear
thickening, even at packing fractions $\phi$ at which shear thickening is
evident in aqueous and brine suspensions \cite{F08}, and no evidence of a
yield stress \cite{BJ09,B10} sufficient to mask it.  When increasing shear
rate is followed, without stopping, by decreasing shear rate there is no
evidence of hysteresis at levels of a few percent.

The absence of shear thickening and hysteresis in oil suspensions of starch
is explained by the ability of oil to wet and spread along metal and glass
surfaces, so that the particulate phase is effectively unconfined, as if it
were in contact with a pool of suspension \cite{F08}.  The sliding of starch
grains along these surfaces is lubricated by films of oil wetting the
surfaces.  The viscosity of oil also makes it more effective in preventing
frictional contacts between grains.

\begin{figure}[h!]
\centering
\subfigure[\ Glycerin/water]{
\includegraphics[width=0.45\textwidth]{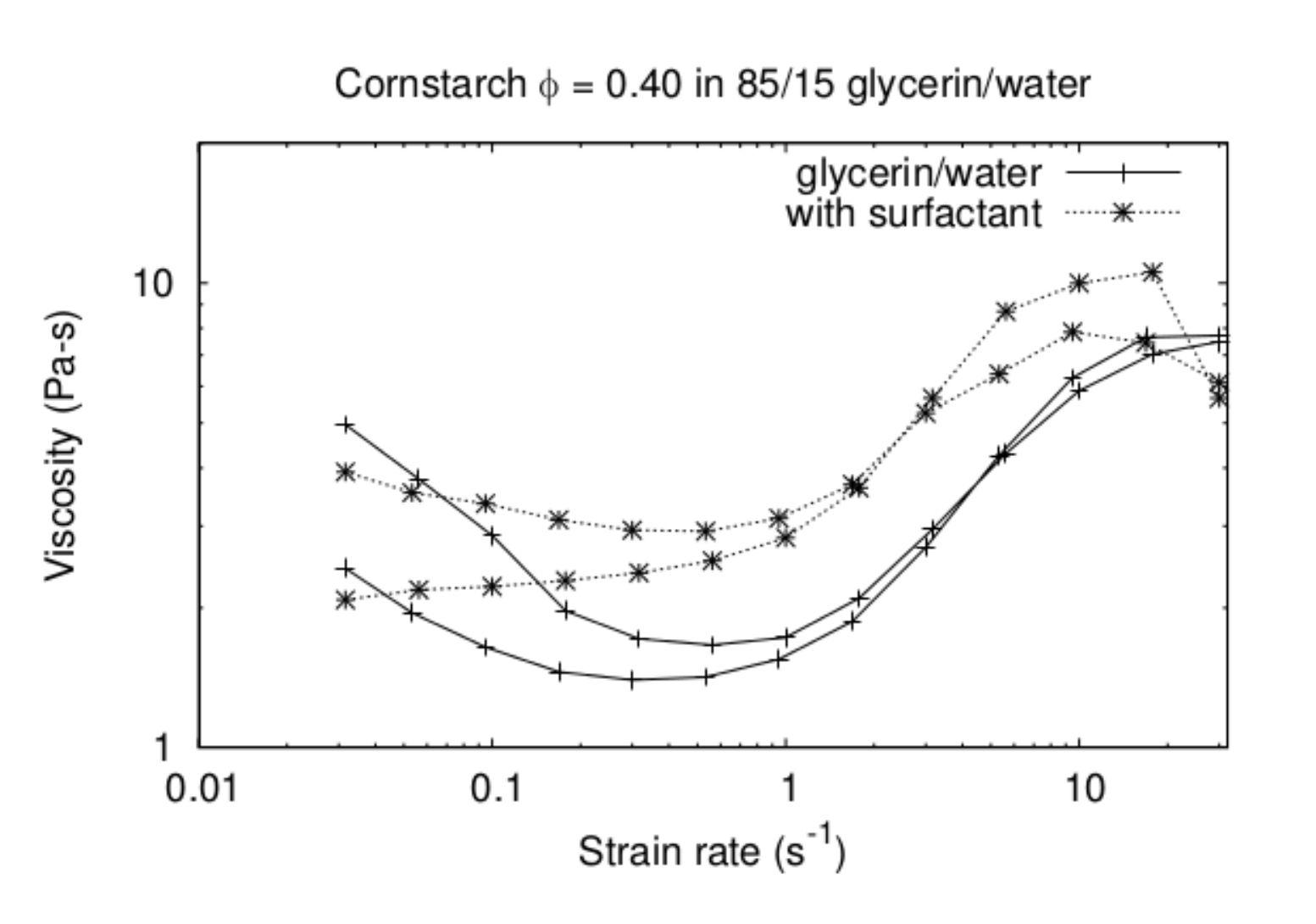}
\label{glycerinwater}}
\subfigure[\ Brine with surfactant]{
\includegraphics[width=0.45\textwidth]{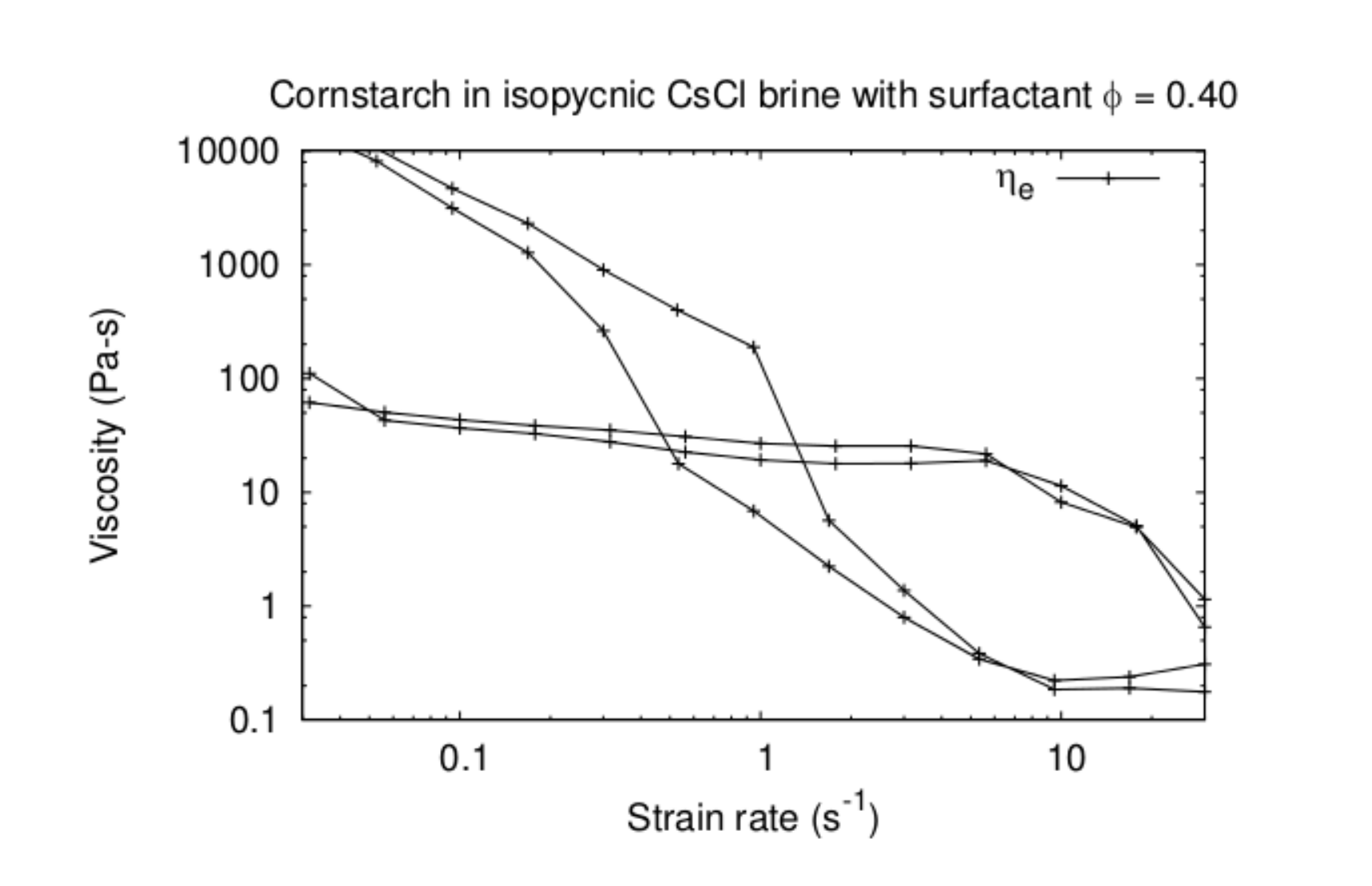}
\label{brinesurf}}
\caption{Rheology of $\phi = 0.40$ suspensions of cornstarch in conical
geometry in: (a) Glycerin/water solution with the same viscosity as olive
oil, with and without miscible surfactant (dish detergent).  (b) Isopycnic
brine with a immiscible surfactant Triton X-100 (two runs).  The
glycerin/water suspension shear thickens, but much less than brine without
surfactant (Fig.~\ref{brine}), continuously and without hysteresis.
The viscous solvent interferes with jamming and DST, plausibly as a
consequence of its greater lubricity.  The surfactant has little effect,
perhaps because the effect of viscosity is sufficient to lubricate contacts,
without regard to surface interactions.  In low viscosity brine the
immiscible surfactant prevents shear thickening entirely, which we suggest
is a consequence of its ability to wet and lubricate confining surfaces.
There is little hysteresis but nominally identical samples have very
different properties, perhaps as a result of differences in the distribution
of buoyant surfactant.}
\label{glycerin/surf}
\end{figure}

A glycerin/water (85\%/15\% by mass) solution is an intermediate between
brine and olive oil as a solvent.  The solvent viscosity equals that of
olive oil, but its molecules are polar and its surface interactions similar
to those of brine.  Sedimentation of concentrated starch in glycerin/water
is slow, as in oil \cite{C00}.  Fig.~\ref{glycerinwater} shows the results
of measurements of the effective viscosity of glycerin/water suspensions of
cornstarch with and without added surfactant.  These may be compared to the
results for brine suspensions in Fig.~\ref{brine}.  There is some shear
thickening, but it is continuous, with little hysteresis, and by only a
factor of about 5, rather than the 2--3 orders of magnitude of suspensions
in brine.  The surfactant does not qualitatively change its properties.  The
greater viscosity (Fig.~\ref{oilfig}) of oil suspensions in the shear
thinning regime may be the result of greater clumping of granules
\cite{B10}, as occurs if they are more readily wetted by the polar solvent
glycerin/water.

The mild shear thickening of glycerin/water suspensions indicates that
viscosity alone is insufficient to prevent shear thickening; wetting is also
required.  Because glycerin/water, like brine, incompletely wets metal and
glass surfaces, contact line forces provide confinement and some shear
thickening is observed.  Its viscosity helps to maintain a lubricating film
of fluid despite incomplete wetting, qualitatively explaining the
observation that its shear thinning is modest in magnitude, gradual in onset
and shows no hysteresis (Fig.~\ref{glycerinwater}).  The normal stress
$\sigma_n$ in the rheometer remains negative, as expected for a wetting
fluid (Fig.~\ref{gap} with $\theta_r = \theta_a = 0$).  Shear dilation
reduces gaps between grains and increases viscous coupling among them,
increasing the suspension viscosity, but is accommodated by sliding and the
grains do not jam.

Fig.~\ref{brinesurf} shows the rheology of a starch 
suspension in isopycnic brine with a small quantity of surfactant (dish
detergent) added.  This surfactant is immiscible in the brine, so that it
forms a third phase that wets and lubricates the confining surfaces in the
same manner as oil.   The particulate phase is effectively unconfined,
precluding jamming and shear thickening.  This result is the opposite of
the effect of a secondary fluid whose addition turns a fluid suspension
into an elastic gel \cite {KW11}.  We attribute the difference to the
surfactant wetting the confining surfaces in our experiments.

\section{Discussion}

In the DST state the grains in cornstarch suspensions in brine jam and shear
between the confining rheometer surfaces with a shear stress that is related
to the confining stress by a coefficient of sliding friction.
Rearrangements of the grains produce a fluctuating shear stress.  In conical
geometry, the entire suspension jams.  Once jammed, the normal stress in the
rheometer maintains the jammed state, even if the angular velocity is
reduced far below the DST threshold.  In parallel plate geometry suspension
near the rheometer axis, where the strain rate is low, does not jam, and may
nucleate an unjamming front that unjams the entire suspension with little
hysteresis as the rotation rate is reduced. 

Surface forces, including both wetting and friction, may explain the DST
threshold strain rate ${\dot \gamma}_c$, hysteresis, and the dependence of
the rheology on the properties of the solvent.  The absence of DST of starch
suspensions in nonpolar solvents may be attributed to their lubricity: if it
is energetically favorable, compared to dry contact, for grains or grains
and confining surfaces to be separated by a thin film of solvent then
lubricated flow prevents jamming.

\begin{acknowledgments}
We thank P.~Bayly for the use of an AR-G2 (TA Instruments) rheometer,
K.~Croat, F. Gyngard, S.~Handley and A.~Nelson for assistance and E.~Brown
for critical reading and advice.  This work was supported in part by
American Chemical Society Petroleum Research Fund Grant \#51987-ND9.
\end{acknowledgments}

\end{document}